\documentclass[aps,superscriptaddress]{revtex4}
\usepackage[T1]{fontenc}
\usepackage[utf8]{inputenc}
\usepackage{mathtools}
\usepackage{graphicx}
\usepackage{caption}
\usepackage{subcaption}
\usepackage{xcolor}
\usepackage{soul}
\usepackage{amsmath,amssymb}
\usepackage[colorlinks=true, pdfstartview=FitV, linkcolor=blue, citecolor=blue, urlcolor=magenta, breaklinks=true]{hyperref}
\usepackage{orcidlink}

\begin{document}

\title{A Bayesian study of quark models in view of recent astrophysical constraints}

\author{Franciele M. da Silva \orcidlink{0000-0003-2568-2901}} 
\email{franmdasilva@gmail.com}

\affiliation{Departamento de F\'isica, CFM - Universidade Federal de Santa Catarina; \\ C.P. 476, CEP 88.040-900, Florian\'opolis, SC, Brazil.}

\author{Adamu Issifu \orcidlink{0000-0002-2843-835X}} 
\email{ai@academico.ufpb.br}

\affiliation{Departamento de F\'isica, CFM - Universidade Federal de Santa Catarina; \\ C.P. 476, CEP 88.040-900, Florian\'opolis, SC, Brazil.}

\author{Luiz L. Lopes \orcidlink{0000-0003-0982-9774}}
\email{llopes@cefetmg.br}

\affiliation{Centro Federal de Educação Tecnológica de Minas Gerais Campus VIII, \\ CEP 37.022-560, Varginha, MG, Brazil}

\author{Luis C. N. Santos \orcidlink{0000-0002-6129-1820}}
\email{luis.santos@ufsc.br}

\affiliation{Departamento de Física, CCEN--Universidade Federal da Paraíba; \\ C.P. 5008, CEP  58.051-970, João Pessoa, PB, Brazil} 
\affiliation{Theoretical Astrophysics, Institute for Astronomy and Astrophysics,
University of Tübingen, \\ 72076 Tübingen, Germany}

\author{D\'ebora P. Menezes \orcidlink{0000-0003-0730-6689}}
\email{debora.p.m@ufsc.br}

\affiliation{Departamento de F\'isica, CFM - Universidade Federal de Santa Catarina; \\ C.P. 476, CEP 88.040-900, Florian\'opolis, SC, Brazil.}

\begin{abstract}

In this work, we perform a comparative analysis between the density-dependent quark model and the vector MIT bag model using Bayesian analysis. We use the equations of state generated by these two models to describe quark stars.
We impose four recent observational astrophysical constraints on both models to determine their model-dependent parameters in an optimized manner assuming that the compact objects observed are composed entirely of self-bound quarks. The restrictions are aimed at producing stars with maximum masses $2 - 2.35$ M$_\odot$ and a mass-radii diagram compatible with the observed pulsars: PSR J0740+6620, PSR J0952-0607, PSR J0030+0451 and the compact object XMMU J173203.3-344518. With this analysis, the parameter dependence of the nuclear equation of state (EoS) of both models is restricted. 

\end{abstract}

\maketitle

\section{Introduction} \label{introduction}

The study of the dense deconfined quark matter phase in nuclear astrophysics is an open problem. There are several theoretical models developed to explore this matter and related phenomena \cite{nambu1961dynamical,fuchs1995density,MITbag}. One of the phenomena that is widely being considered these days is the probing of the inner core of the neutron star (NS). The recently operating Neutron Star Interior Composition Explorer (NICER) \cite{NANOGrav:2019jur, Fonseca:2021wxt, Riley:2021pdl} and the gravitational wave laser interferometer (Advanced LIGO, Virgo, and KAGRA) \cite{LIGOScientific:2017vwq, LIGOScientific:2016aoc, LIGOScientific:2018cki, LIGOScientific:2018hze, Miller:2023qyw} have started providing results that constrain some known properties of the NSs including their deformability, maximum masses and radii. The NSs are remnants of gravitationally collapsed supernovae that lead to the formation of the densest and smallest observed compact objects in the universe. Their densities are several times higher than the density of ordinary nuclear matter \cite{Menezes:2021jmw, Baym:2017whm}. The hadronic matter forming the star can undergo a phase transition to quark matter in the core of the star where matter is expected to be highly dense creating a hybrid star \cite{Lopes:2021jpm, 10.1093/mnras/stad2509, Maruyama:2007ey, Paoli:2010kc}. Hypothetically, NSs could also be formed through self-bound deconfined quarks making up the entire star, which is effectively a quark star, also known as strange star \cite{Glendenning1997, Witten:1984rs, PhysRevD.4.1601,Linares2006}. Indeed, the physical consequences of deconfined quark matter in NS/protoneutron stars and core-collapse supernovas have a long-standing history in observational astrophysics \cite{Alford:2006vz, PhysRevD.76.123015, Weissenborn:2011qu, Pons:2001ar, Schertler:2000xq}.

The EoS mainly consists of the pressure as a function of the energy density and carries information on the inner dynamics and composition of the NS. Through the EoS we can determine the macroscopic nature of the star via the Tolman–Oppenheimer–Volkoff (TOV) equations \cite{PhysRev.55.374}. The EoS also serves as the basis for several astrophysical simulations of compact objects so, several investigations are done to improve its accuracy \cite{CompOSECoreTeam:2022ddl, Dutra:2014qga}. Additionally, experimental nuclear physics provides valuable data required to benchmark theories of dense matter EoS in NS \cite{PhysRevC.71.024312}. The asymptotic freedom behavior of the quantum chromodynamics (QCD) theory that characterizes the phase transition of strongly interacting matter from hadron phase to deconfined quark matter phase is an important subject in the study of the QCD phase diagram. The deconfined quark matter phenomena are expected to also occur at the higher baryon density region of NS matter \cite{Fukushima:2010bq, Gross:1973id}.

Proper understanding of the EoSs for high-density, cold quark matter plays a significant role in constraining the characteristics of strongly interacting matter believed to exist in the core of NSs. The EoS at high density for cold quark matter is considered a robust constraint \cite{Komoltsev:2021jzg, Gorda:2023mkk} when constructing the NS EoS at low densities \cite{Annala:2017llu, Most:2018hfd, Altiparmak:2022bke, Annala:2021gom} as well. Aside from its phenomenological applications, the high-density cold quark matter EoS has enormous theoretical applications due to its associated rich physics, including dynamical screening of long-wavelength chromoeletric and chromomagnetic fields. Such screening behaviors are also expected to occur in a high-temperature quark-gluon plasma regime \cite{Bazavov:2020teh, Karsch:1987pv}.

Ever since Bodmer and Witten hypothesized that stable quark matter should necessarily contain strange quark matter (SQM) \cite{PhysRevD.4.1601, Witten:1984rs}, it has been assumed that quark stars consist of SQM \cite{Glendenning1997}. The Bodmer-Witten conjecture states that the energy per baryon of SQM at zero pressure must be less than the one observed in the infinite baryonic matter:
\begin{equation}
    \varepsilon(p=0)_{SQM} = E/A \leq 930~\mbox{MeV}. \label{l930}
\end{equation}
At the same time, the non-strange quark matter (NSQM) still needs to have an energy per baryon higher than the nonstrange infinite baryonic matter; otherwise, protons and neutrons would decay into $u$ and $d$ quarks:
\begin{equation}
    \varepsilon(p=0)_{NSQM} = E/A > 930~\mbox{MeV}. \label{g930}
\end{equation}
Therefore, Equations (\ref{l930}) and (\ref{g930}) must be simultaneously satisfied if a model is used to describe quark stars. The same is not true in the construction of hybrid stars, in which just their cores are constituted of deconfined quarks.

In this work, we perform a comparative analysis of the density-dependent quark model (DDQM) \cite{Wu:2008tc, PhysRevC.61.015201, Xia:2014zaa, Backes:2020fyw, Wen:2005uf} and the vector MIT bag model \cite{Cierniak:2018aet, MITbag, Lopes_2021,Lopes_2021b} using Bayesian analysis. The Bayesian approach is a very useful tool for inferring the probability distribution of a set of model parameters based on a set of measured data. In the context of nuclear physics and astrophysics, the Bayesian analysis can be used to optimize a set of EoS parameters given astrophysical observations \cite{zhu2023bayesian,takatsy2023neutron,de2023bayesian}, as well as nuclear matter properties \cite{wesolowski2016bayesian,traversi2020bayesian,malik2023spanning,jiang2023bayesian}. We rely on recent observational astrophysical data that leads to robust constraints on the NS EoS and its internal composition assuming that the NSs are formed entirely by self-bound quarks in a deconfined state. The constraints include measurable global properties of the NS such as the mass and radii of PSR J0740+6620 \cite{riley2021nicer}, PSR J0952-0607 \cite{romani2022psr}, PSR J0030+0451 \cite{riley2019nicer}  and XMMU J173203.3-344518 \cite{doroshenko2022strangely}. Notice that in the present work, we restrict ourselves to quark (strange) stars only.

With regards to the DDQM, we determine optimal values of the low-density parameter, $D$, which is related to the linear confining properties of the quarks, and the higher-density parameter, $C$, which in turn is associated with the leading order perturbative interaction term in the QCD theory. These are model-dependent parameters that are usually determined in the model framework. We determine these parameters using Bayesian analysis that conforms with each observed star listed above. On the other hand, with the vector MIT bag model, we determine the model-dependent parameters such as the bag constant, $B$, related to the vacuum pressure, the strength of the vector coupling, $G_V$, responsible for the quark-quark repulsion, and the self-coupling channel, $b_4$, which mimics the Dirac sea contribution and it is important to soften the EoS at very high densities.

We determine each of these constants that satisfy a particular observed star for each of the models and compare their properties. The NS properties that we consider for comparison and analysis are the EoS, sound velocity, $c_s$, tidal deformability, $\Lambda$, mass, $M$, and radius, $R$ and the adiabatic index, $\Gamma$.

This work is organized as follows: In Section \ref{sec_ddqm} we introduce the DDQM and discuss the fundamental relations that govern it. In Section \ref{sec_MIT} we introduce the vector MIT bag model and discuss its fundamental properties. A brief discussion of the mass-radius constraints and their associated expressions was presented in Section \ref{sec_mr}. In Section \ref{sec_bayes} we present the general overview of the Bayesian analysis and the corner diagrams for the four different cases (Figures \ref{fig1}, \ref{fig2}, \ref{fig3} and \ref{fig4}) considered in this work for both models. The analysis of the stability window of the DDQM was also discussed briefly in Subsection \ref{window}. The results and analysis of the study are presented in Section \ref{sec_res} and the final findings in Section \ref{conclusions}.

\section{Density-dependent quark model} \label{sec_ddqm}

The density-dependent quark model (DDQM) is a model that describes the SQM and incorporates the interaction between the quarks through a dependency of the mass on the density. In this work, we are using the model given in ref. \cite{xia2014thermodynamic}: 
\begin{equation}
    m_i = m_{i0} + \frac{D}{n^{1/3}} + C n^{1/3},
\end{equation}
where  $m_{i0} (i  = u,\, d,\, s)$ is the current quark mass, $n$ is the baryon number density, and $C$ and $D$ are the parameters of this model. A possible problem with the introduction of a density dependency is that it can lead to thermodynamic inconsistencies. A way to avoid these inconsistencies is the inclusion of an effective chemical potential $\mu^*_i$, and in this way, we can describe the system by a free-energy density $f$ of a free particle system with masses $m_i (n)$ and effective chemical potentials $\mu^*_i$
\begin{equation}
    f = \Omega_0 \left( \{ \mu^*_i \}, \{ m_i \} \right) + \sum_i \mu^*_i n_i,
\end{equation}
where $\Omega_0$ is the thermodynamic potential density of the free quarks, given by the following expression
\begin{equation}
    \Omega_0 = - \sum_i \frac{\gamma_i}{24 \pi^2} \left[ \mu^*_i \nu_i \left( \nu_i^2 -\frac{3}{2} m_i^2 \right) +\frac{3}{2} m_i^4 \ln{\frac{\mu^*_i +\nu_i}{m_i}}\right], 
\end{equation}
with $\gamma_i=6~ (3~ {\rm colors} \times 2~ {\rm  spins})$ is the degeneracy factor. The Fermi momenta is given in terms of the effective chemical potentials $\mu^*_i$:
\begin{equation}
    \nu_i = \sqrt{\mu^{*2}_i - m_i^2},
\end{equation}
so that the particle number density $n_i$ can be written as 
\begin{equation}
    n_i = \frac{\gamma_i}{6 \pi^2} (\mu^{*2}_i - m_i^2)^{3/2} = \frac{\gamma_i \nu_i^3}{6 \pi^2}
\end{equation}
and the chemical potential $\mu_i$ and the effective chemical potential are related through the relation
\begin{equation}
    \mu_i = \mu_i^* - \mu_I.
\end{equation}
The $\beta$-equilibrium condition can be rewritten in terms of $\mu^*_i$ as:
\begin{equation}
    \mu_u^* + \mu_e = \mu_d^* = \mu_s^*.
\end{equation}
To construct the EoS we also take into consideration the usual charge neutrality condition
 \begin{equation}
    \frac{2}{3} n_u - \frac{1}{3} n_d - \frac{1}{3} n_s - n_e = 0,
\end{equation}
and the baryon number conservation
\begin{equation}
    n = \frac{1}{3} (n_u +n_d +n_s).
\end{equation}

This way, the energy density $\varepsilon$ of the system is given by
\begin{equation}
    \varepsilon = \Omega_0 - \sum_i \mu_i^* \frac{\partial \Omega_0}{\partial \mu_i^*},
\end{equation}
and the pressure $p$ by
\begin{equation}
    p = -\Omega_0 + \sum_{i,j} \frac{\partial \Omega_0}{\partial m_j} n_i \frac{\partial m_j}{\partial n_i}.
\end{equation}

\section{Vector MIT bag model} \label{sec_MIT}

The vector MIT bag model is an extension of the original MIT bag model~\cite{MITbag} that incorporates some
features of the quantum Hadrodynamics (QHD)~\cite{Serot_1992}. In its original form, the MIT bag model considers that each baryon is composed of three non-interacting quarks inside a bag. The
bag, in turn, corresponds to an infinite potential that confines the quarks. As a consequence, the quarks are free inside the bag and are forbidden to reach its exterior. All the information about the strong force relies on the bag pressure value, which mimics the vacuum pressure.

In the vector MIT bag model, the quarks are still confined inside the bag, but now they interact with each other through a vector meson exchange. This vector meson plays a role analog to the $\omega$ meson of the QHD~\cite{Serot_1992}. Moreover, the contribution of the Dirac sea can be taken into account through a self-interaction of the vector meson~\cite{furnstahl1997vacuum}. The Lagrangian of the vector MIT bag model, therefore, consists of the Lagrangian of the original MIT, plus the Yukawa-type Lagrangian of the vector field exchange, plus the Dirac sea contribution. We must also add the mesonic mass term to maintain the thermodynamic consistency. It then reads~\cite{Lopes_2021,Lopes_2021b}:

\begin{equation}
    \mathcal{L} =  \mathcal{L}_{MIT} +  \mathcal{L}_V + \mathcal{L}_{DIRAC}, \label{l1}
\end{equation}
where
\begin{equation}
\mathcal{L}_{MIT} = \sum_{i}\{ \bar{\psi}_i  [ i\gamma^{\mu} \partial_\mu - m_i ]\psi_i - B \}\Theta(\bar{\psi_i}\psi_i), \label{e1}
\end{equation}  

\begin{equation}
 \mathcal{L}_V = \sum_{i}\{\bar{\psi}_i g_{iV}(\gamma^\mu V_{\mu})\psi_i - \frac{1}{2}m_V^2 V^\mu V_\mu \} \Theta(\bar{\psi}_i\psi_i) ,\label{Ne9},
\end{equation}

\begin{equation}
 \mathcal{L}_{DIRAC} = b_4\frac{(g^2V_\mu V^\mu)^2}{4} ,\label{e12}
\end{equation}
where $\psi_i$ is the Dirac quark field, $B$ is the constant vacuum pressure, $m_V$ is the mass of the $V_0$ mesonic field, $g_{iV}$ is the coupling constant of the quark $i$ with the meson $V_0$, and $g$ = $g_{uV}$. The  $\Theta(\bar{\psi}_i\psi_i)$ is the Heaviside step function included to assure that the quarks exist only confined to the bag

Applying mean-field approximation (MFA)~\cite{Serot_1992}, and the Euler-Lagrange equations, we can obtain the energy eigenvalue for the  quark fields, and the equation of motion for the mesonic $V_0$ field:

\begin{equation}
 E_i =  \sqrt{m_i^2 + \nu_i^2} +g_{iV}V_0 \label{eigen}   
\end{equation}

\begin{equation}
gV_0  + \bigg ( \frac{g}{m_v} \bigg)^2 \bigg (  b_4 (gV_0)^3 \bigg )= \nonumber 
 \bigg (\frac{g}{m_v} \bigg ) \sum_{i} \bigg (\frac{g_{iV}}{m_v} \bigg )n_i . \nonumber \\ \label{V0}
\end{equation}

To construct an EoS in MFA,  we now consider the Fermi-Dirac distribution of the quarks, and the Hamiltonian of the vector field, and the bag pressure value, $\mathcal{H} = -\langle \mathcal{L} \rangle$. We obtain:

\begin{equation}
\varepsilon_i = \frac{\gamma_i}{2\pi^2}\int_0^{\nu_f} E_i ~\nu^2 d\nu,\label{nl4}
\end{equation}

\begin{equation}
\varepsilon =  \sum_i \varepsilon_i + B - \frac{1}{2}m_V^2V_0^2  - b_4\frac{(g^2V_0^2)^2}{4}. \label{nl5}
\end{equation}
Now we define $G_V~\equiv~(g/m_V)^2$ and $X_V~\equiv~(g_{sV}/g_{uV})$.  The $X_V$ is then taken as $X_V$ = 0.4, once its value was calculated based on symmetry group arguments (see reference \cite{Lopes_2021} for additional details). Finally, the pressure is easily obtained by thermodynamic relations: $p =
\sum_i \mu_i n_i - \varepsilon$.

\section{Constraints on mass-radius relations} \label{sec_mr}

In this Section, we discuss the recent astrophysical observations and their connections with equilibrium properties associated to a specific EoS. By considering a spherically symmetric body, we are interested in determining the mass and the radius associated. Due to the high-density matter of compact objects such as quark stars, it is assumed that the correct equilibrium properties can be obtained by solving the TOV  equations~\cite{PhysRev.55.374}:

\begin{align}
    \frac{dp(r)}{dr}&=-[\varepsilon(r) + p(r)]\frac{M(r)+4\pi r^3 p(r)}{r^2-2M(r)r}, \label{eq1}\\
    \frac{dM(r)}{dr}&=4\pi r^2 \varepsilon(r), \label{eq2}
\end{align}
where $M(r)$ is the gravitational mass associated with a spherically symmetric compact object with radius $R$, $p(r)$, and $\varepsilon(r)$ { are the pressure and the energy density respectively, and we have used natural units such that $G=c=1$. }
For realistic models of compact objects, Equations (\ref{eq1}) and (\ref{eq2}) are usually solved by using numerical techniques. In this regard, we consider a compact star with central energy density $\varepsilon(r=0)=\varepsilon_c$ and total mass $M$ obtained using the boundary condition $p(R)=0$ where $R$ is radius of the star. The next step in solving the hydrostatic equilibrium equations consists of determining the connection between energy density and pressure. This relation depends on the model we are using to construct the EoS associated with the astrophysical object. In this study, we utilize the models introduced in Sections \ref{sec_ddqm} and \ref{sec_MIT} in order to obtain mass-radius relations and use recent observational constraints to restrict the model-dependent parameters. 

Recent measurements by the NICER mission are advancing our knowledge of the constraints that should be considered for a given EoS of dense matter performing strict limits on the radius. With information only about the mass and radius of a neutron star, its exact internal structure remains uncertain. It is expected that multi-messenger astronomy, such as gravitational waves, can be used in order to provide signatures of the composition of the star. An example of an effect that can be detected is the signature of some quasi-normal modes (QNMs) associated with tidal forces between neutron stars merging into gravitational waves \cite{qnm}. 

Regarding the analysis carried out in this paper, the millisecond pulsar PSR J0740+6620 is an interesting system that orbits with a binary companion. Due to a favorable inclination, the Shapiro time delay was used to measure the mass of this source with remarkable precision, making this one of the most well-constrained massive neutron stars known \cite{cromartie2020relativistic,fonseca2021refined,riley2021nicer}. The timing of PSR J0740+6620 made with data from the North American Nanohertz Observatory for Gravitational Waves (NANOGrav) in combination with observations from the Green Bank Telescope led to a mass estimate of $2.14_{-0.09}^{+0.10}$ M$_{\odot}$ ($68.3 \%$ credibility interval) \cite{cromartie2020relativistic}. Continued timing observations of this massive millisecond pulsar allowed the improvement of this estimate, leading to a mass of $2.08_{-0.07}^{+0.07}$ M$_{\odot}$ ($68.3 \%$ credibility interval) \cite{fonseca2021refined}.

\section{Bayesian analysis} \label{sec_bayes}

\begin{table}[t]
    \centering
        \caption{Mass and radius of the compact stars used as constraints in this work.}
    \begin{ruledtabular}
    \begin{tabular}{l c c}
         \textbf{Star}  &  \textbf{Mass}  &  \textbf{Radius}  \\
         \hline
         PSR J0952-0607 \cite{romani2022psr}  &  $2.35 \pm 0.17$ M$_{\odot}$  &  ---  \\

         PSR J0740+6620 \cite{riley2021nicer}  &   $2.072_{-0.066}^{+0.067}$ M$_{\odot}$  &  $12.39_{-0.98}^{+1.30}$ km  \\

         PSR J0030+0451 \cite{riley2019nicer}  &  $1.34_{-0.16}^{+0.15}$ M$_{\odot}$  &  $12.71_{-1.19}^{+1.14}$ km  \\ 

         XMMU J173203.3-344518 \cite{doroshenko2022strangely}  &  $0.77_{-0.17}^{+0.20}$ M$_{\odot}$  &  $10.4_{-0.78}^{+0.86}$ km 
    \end{tabular}
    \end{ruledtabular}
    \label{tab1}
\end{table}

\begin{figure}[t]
     \centering
     \begin{subfigure}{0.49\textwidth}
         \centering
         \includegraphics[width=\textwidth]{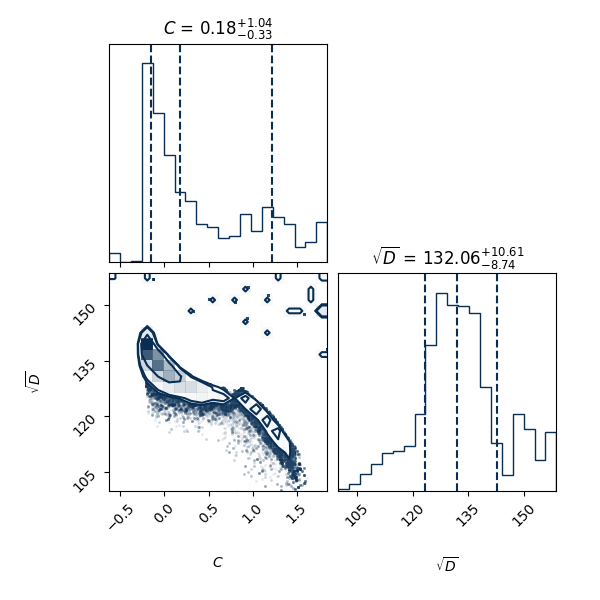}
     \end{subfigure}
     \begin{subfigure}{0.49\textwidth}
         \centering
         \includegraphics[width=\textwidth]{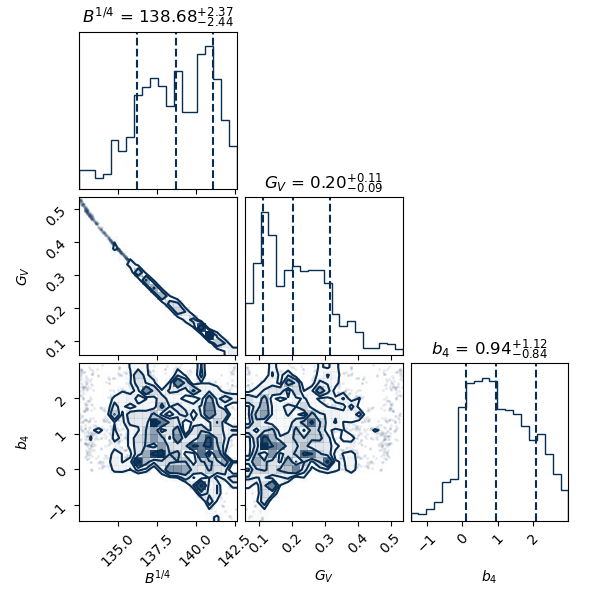}
     \end{subfigure}
        \caption{Corner plot showing the posterior distributions of the parameters of DDQM model on the left, and the parameters of the vector MIT bag model on the right for Case I. The dark to light contours represent the $1 \sigma$, $2 \sigma$ and $3 \sigma$, respectively. The dashed vertical lines in the histograms represent the 0.16, 0.5, and 0.84 quantiles.}
        \label{fig1}
\end{figure} 

\begin{figure}[t]
     \centering
     \begin{subfigure}{0.49\textwidth}
         \centering
         \includegraphics[width=\textwidth]{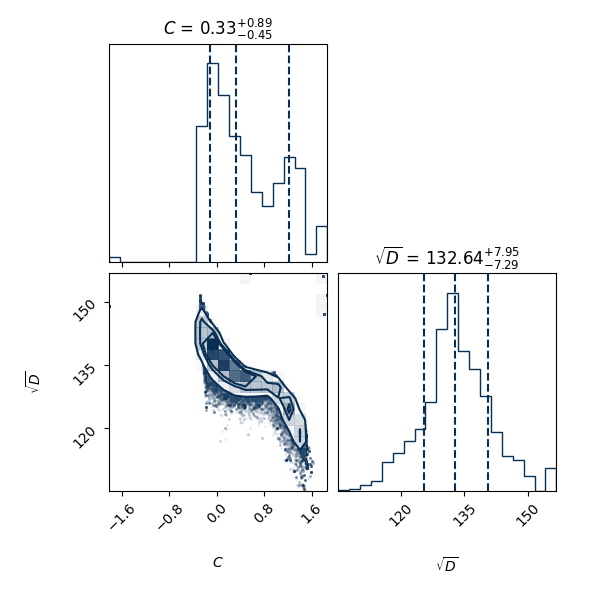}
     \end{subfigure}
     \begin{subfigure}{0.49\textwidth}
         \centering
         \includegraphics[width=\textwidth]{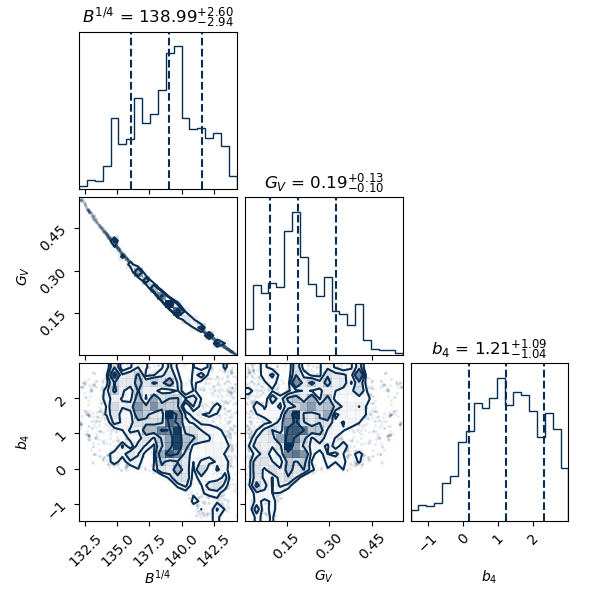}
     \end{subfigure}
        \caption{Corner plot showing the posterior distributions of the parameters of DDQM model on the left, and the parameters of the vector MIT bag model on the right for Case II. The dark to light contours represent the $1 \sigma$, $2 \sigma$ and $3 \sigma$, respectively. The dashed vertical lines in the histograms represent the 0.16, 0.5, and 0.84 quantiles.}
        \label{fig2}
\end{figure}

\begin{figure}[t]
     \centering
     \begin{subfigure}{0.49\textwidth}
         \centering
         \includegraphics[width=\textwidth]{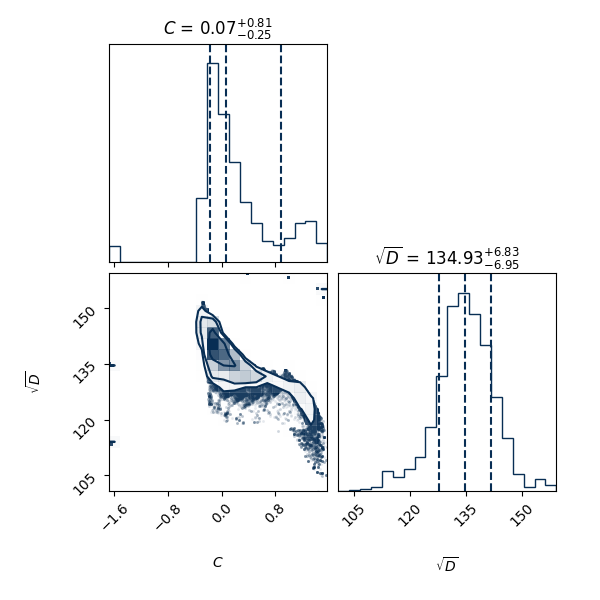}
     \end{subfigure}
     \begin{subfigure}{0.49\textwidth}
         \centering
         \includegraphics[width=\textwidth]{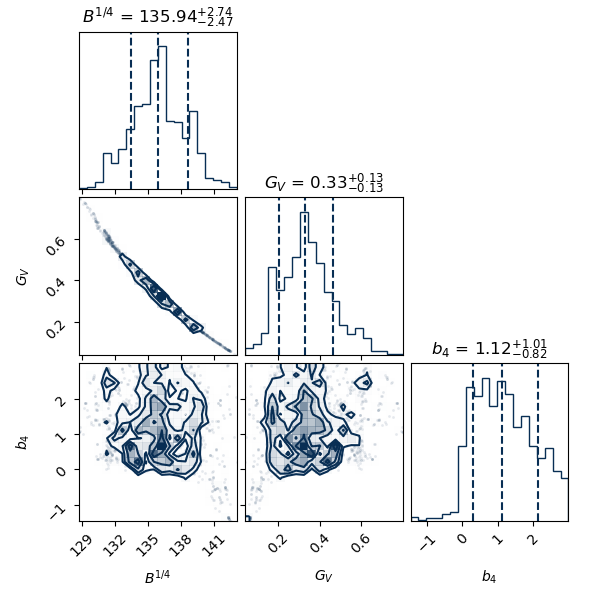}
     \end{subfigure}
        \caption{Corner plot showing the posterior distributions of the parameters of DDQM model on the left, and the parameters of the vector MIT bag model on the right for Case III. The dark to light contours represent the $1 \sigma$, $2 \sigma$ and $3 \sigma$, respectively. The dashed vertical lines in the histograms represent the 0.16, 0.5, and 0.84 quantiles.}
         \label{fig3}
\end{figure}

\begin{figure}[t]
     \centering
     \begin{subfigure}{0.49\textwidth}
         \centering
         \includegraphics[width=\textwidth]{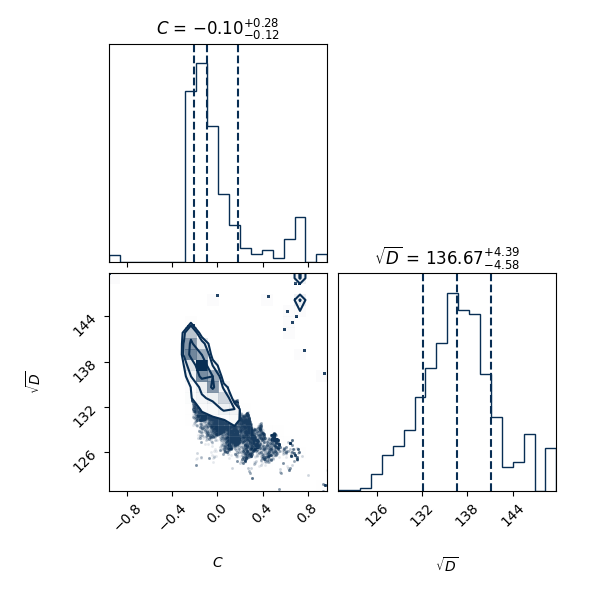}
     \end{subfigure}
     \begin{subfigure}{0.49\textwidth}
         \centering
         \includegraphics[width=\textwidth]{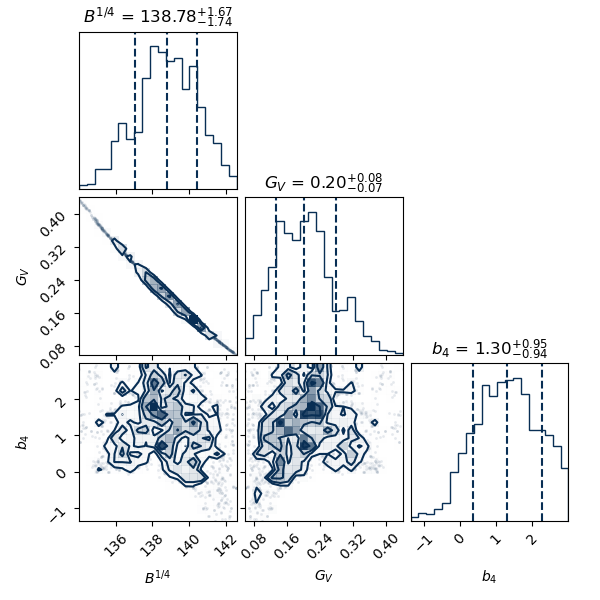}
     \end{subfigure}
        \caption{Corner plot showing the posterior distributions of the parameters of DDQM model on the left, and the parameters of the vector MIT bag model on the right for Case IV. The dark to light contours represent the $1 \sigma$, $2 \sigma$ and $3 \sigma$, respectively. The dashed vertical lines in the histograms represent the 0.16, 0.5, and 0.84 quantiles.}
        \label{fig4}
\end{figure} 

\begin{figure}[t]
    \centering
    \includegraphics{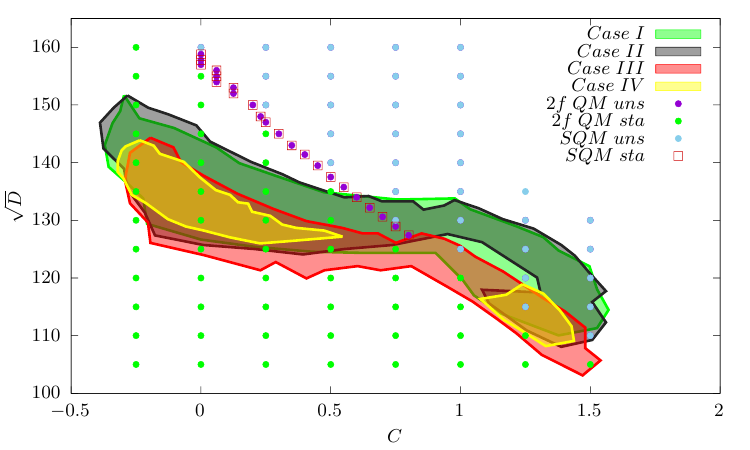}
    \caption{The stability window for the DDQM model.}
    \label{fig5}
\end{figure}

In this work, we have optimized the EoS parameters for the constraints investigated by means of a Bayesian analysis. In this approach given an EoS model $\mathcal{M}$ and a set of EoS parameters $\boldsymbol{\theta}$, we have an estimate of the region in which the values of the parameters lie. This information is given probabilistically by the prior distribution $\mathcal{P}(\boldsymbol{\theta}, \mathcal{M})$. We can increase our information about $\boldsymbol{\theta}$ if we observe what is the chance of measuring a specific data $\mathfrak{D}$ when we consider a specific value of $\boldsymbol{\theta}$, this increase of information is accounted by the likelihood function $\mathcal{L}(\mathfrak{D}|\boldsymbol{\theta},\mathcal{M})$. So, with Bayes's theorem, we can obtain a posterior probability $\mathcal{P}(\boldsymbol{\theta}|\mathfrak{D},\mathcal{M})$ for $\boldsymbol{\theta}$ given a data set $\mathfrak{D}$
\begin{equation}
    \mathcal{P}(\boldsymbol{\theta}|\mathfrak{D},\mathcal{M}) \propto \mathcal{L}(\mathfrak{D}|\boldsymbol{\theta},\mathcal{M}) \mathcal{P}(\boldsymbol{\theta}, \mathcal{M}).
\end{equation}
As already explained, we test two models $\mathcal{M}$: DDQM and vector MIT bag. For the DDQM model $\boldsymbol{\theta}=\{C,~ \sqrt{D}\}$, and for the vector MIT bag model, $\boldsymbol{\theta}=\{B^{1/4},~G_V,~b_4\}$, and the data set $\mathfrak{D}$ are the masses and radii of four compact stars present in Table \ref{tab1}. The posteriors $\mathcal{P}(\boldsymbol{\theta}|\mathfrak{D},\mathcal{M})$ were obtained using the \textit{emcee} package \cite{foreman2013emcee}, which uses the Goodman and Weare's Affine Invariant Markov chain Monte Carlo (MCMC) \cite{goodman2010ensemble} method for sampling the posterior probability density.

In all the cases analyzed we have used a uniform prior and the domain of our priors was based on the previous works on the EoS we are testing. For the DDQM model, the prior domain was taken from references \cite{Wen:2005uf,Backes:2020fyw}, and the range in which the parameters $C$ and $D$ were tested were $C=\{-2, 2\}$ and $\sqrt{D}= \{0~ {\rm MeV}, 300~ {\rm MeV} \}$. As for the vector MIT bag model, we based our choices on \cite{Lopes_2021,lopes2022nature}, and the range in which parameters $B^{1/4}$, $G_V$ and $b_4$ where tested were  $B^{1/4}=\{130~ {\rm MeV}, 170~ {\rm MeV}\}$, $G_V=\{0~ {\rm fm}^2, 2~ {\rm fm}^2\}$  and $b_4= \{-2, 2 \}$.

In Table \ref{tab1} we show the compact stars that we choose as constraints in this work and their respective masses and radii. The first star on the list is the ``black widow'' PSR J0952-0607, which is the fastest known spinning pulsar in the Milky Way, with a frequency of $707$ Hz, and also has the largest mass measured with good accuracy found so far \cite{romani2022psr}. The second star on our table is the massive millisecond pulsar PSR J0740+6620, which had its mass and radius estimated from data of the NICER collaboration \cite{riley2021nicer}. Our third star is the millisecond pulsar PSR J0030+0451 which also had its mass and radius estimated from NICER \cite{riley2019nicer}, and can be used to constrain the radius of the canonical neutron star ($1.4$ M$_\odot$). The last star in our table is the very low-mass compact star XMMU J173203.3-344518 which is inside the supernova remnant HESS J1731-347 and, due to its low mass, was suggested to be a quark star \cite{horvath2023light,Lopes2023MNRAS}. However, we would like to point out that the mass-radius measurement of this object has been contested by some groups, see for example \cite{alford2023central}.

For all the data present in Table \ref{tab1} we have assumed a Gaussian likelihood function
\begin{equation}
    \mathcal{L}(\mathfrak{D}|\boldsymbol{\theta},\mathcal{M}) = \prod_i \frac{1}{\sqrt{2 \pi \sigma_i^2}} e^{-\frac{1}{2}\left(\frac{d_i-m_i(\theta)}{\sigma_i}\right)^2}
\end{equation}
were $d_i$ and $m_i$ are the data and corresponding model values, and the uncertainty $\sigma_i$ for each case is given by the highest data uncertainty, for example, for $r_i=10.4_{-0.78}^{+0.86}$ km we take $\sigma_i=0.86$ km.
In all cases analyzed we assumed causal limits for the mass and radius of the compact stars, in the case of the mass we considered the limit M$_{max} < 3.2$ M$_{\odot}$ \cite{rhoades1974maximum} and, for the radius we used the limit $R > 3 M $. 
In this way, if an EoS produces stars outside these limits we attribute $\mathcal{L}(\mathfrak{D}|\boldsymbol{\theta},\mathcal{M})=0$ for the parameters $\boldsymbol{\theta}$ that lead to this result. We performed the stability window calculations for both models using the conditions from Equations (\ref{l930}) and (\ref{g930}), and considering the following quark masses: $m_u=2.16$ MeV, $m_d=4.67$ Mev and $m_s=93.4$ MeV \cite{pdg}.
For the analysis of the vector MIT bag model, we were able to take the stability window into account in the inference, this way, all the points shown in Figures \ref{fig1} to \ref{fig4} for this model are inside the stability window. 
Values previously obtained can be found in Table 4 of ref. \cite{Lopes_2021}.
As for the DDQM model, we show its stability window analysis in the next subsection.

\begin{itemize}

\item {\bf CASE I:} In Case I we have tested if the quark matter EoS studied here can describe two compact stars with high masses. One of them is the pulsar PSR J0952-0607 \cite{romani2022psr} and the other one is the pulsar PSR J0740+6620 \cite{riley2021nicer}. We want to find EoSs that lead to mass-radius curves that have the masses and the radii of PSR J0952-0607 and  PSR J0740+6620 in some point of the curve. We do not demand a specific value for M$_{max}$, only that EoSs that lead to M$_{max} \leq 2.18$ M$_{\odot}$, which corresponds to the mass of PSR J0952-0607 less its error margin, are forbidden. This way, if the maximum mass is outside the region $2.18$ M$_{\odot} \leq M_{max} \leq 3.2$ M$_{\odot}$ we associate $\mathcal{L}(\mathfrak{D}|\boldsymbol{\theta},\mathcal{M})=0$ to the parameters that lead to this result. In Figure \ref{fig1}, we show the corner plots \cite{corner} that resulted from the Bayesian analysis for the DDQM model on the left panel and for the vector MIT bag model on the right panel. Based on the result for the MIT case we selected the point $\{B^{1/4}=139.79$ MeV, $G_V=0.159$ fm$^2, b_4=1.69 \}$ to be analyzed. As for the DDQM model, we discuss the points chosen to be analyzed in Subsection \ref{window}. 
 
\item {\bf CASE II:} In Case II we have checked if the EoS studied here can describe data from NICER. In this case, we consider a canonical star (PSR J0030+0451) and another star with a mass around $2$ M$_{\odot}$ (PSR J0740+6620) which, according to data from NICER have approximately the same radius. Hence, we want EoSs that lead to mass-radius diagrams with the masses and radii of PSR J0030+0451 and PSR J0740+6620 at some point of their curves and, we disregard EoSs that lead to maximum mass outside the region $2.005$ M$_{\odot} \leq M_{max} \leq 3.2$ M$_{\odot}$. In Figure \ref{fig2}, we can see the plots for this case and, based on these results we selected the point $\{B^{1/4}=140.90$ MeV, $G_V=0.116$ fm$^2, b_4=0.72 \}$ for the MIT bag model to be analyzed.

\item {\bf CASE III:} In Case III we have investigated if the EoS we are studying can describe two compact stars with small masses. One of these stars is XMMU J173203.3-344518 \cite{horvath2023light} and the other one is the canonical star from the previous case. In this case, we assume that $2$ M$_{\odot} \leq M_{max} \leq 3.2$ M$_{\odot}$. From Figure \ref{fig3}, which was obtained from the Bayesian analysis for this case, we selected the point $\{B^{1/4}=135.28$ MeV, $G_V=0.366$ fm$^2, b_4=1.90 \}$ of the MIT bag case to be analyzed.

\item {\bf Case IV:}
Lastly, we want to verify if the EoSs we are studying can describe all of the stars of previous cases at the same time. For Case IV, we assume the same limit for M$_{max}$ as for Case I and, based on the right panel of Figure \ref{fig4}, we chose the point $\{B^{1/4}=137.96$ MeV, $G_V=0.235$ fm$^2, b_4=1.63 \}$ to be analyzed.
\end{itemize}

\subsection{DDQM stability window} \label{window}

In Figure \ref{fig5}, 
we show the stability window for the DDQM EoS. We have the values of the parameter $C$ in the x-axis and the values of $\sqrt{D}$ in MeV in the y-axis. The lower region with green dots is a forbidden region because it represents the region where the quark matter (QM), composed of quarks $u$ and $d$, would be stable. The upper region with blue dots is the region where the SQM is unstable and hence, the matter with EoS obtained from values of $C$ and $\sqrt{D}$ in this region cannot form quark stars. Lastly, the region where we have at the same time purple dots, which represent unstable $u-d$ quark matter, and red squares, which represent stable SQM, is the region we are interested in since the EoSs obtained from values of  $C$ and $\sqrt{D}$ in this region fulfill the requirements to form strange stars. To follow the discussion below, please, refer to Figures \ref{fig1}, \ref{fig2}, \ref{fig3} and \ref{fig4} explained in Section \ref{sec_bayes}.
The shaded areas in Figure \ref{fig5} represent each of the probability distributions we can observe in Figures \ref{fig1} to \ref{fig4} for the DDQM EoS. We can observe that most parts of the distributions lie in the regions where quark stars are not possible. This way for the DDQM model, although we have chosen points to study based on Bayesian analysis, it is not possible to associate each of the points with Cases I to IV, as we did with the vector MIT bag model. Based on Figure \ref{fig5} we have chosen 4 points to analyze, one of them $\{0.5,137.5~ {\rm MeV}\}$ is outside the shaded regions and, as can be seen in Figure \ref{fig6} does not satisfy the requirement M$_{max} \geq 2 {\rm M}_{\odot}$. Some of the main properties of each point can be seen in Table \ref{tabdd}. 

\section{Results and Analysis} \label{sec_res}

\begin{figure}[t]
     \centering
     \begin{subfigure}{0.49\textwidth}
         \centering
         \includegraphics[width=\textwidth]{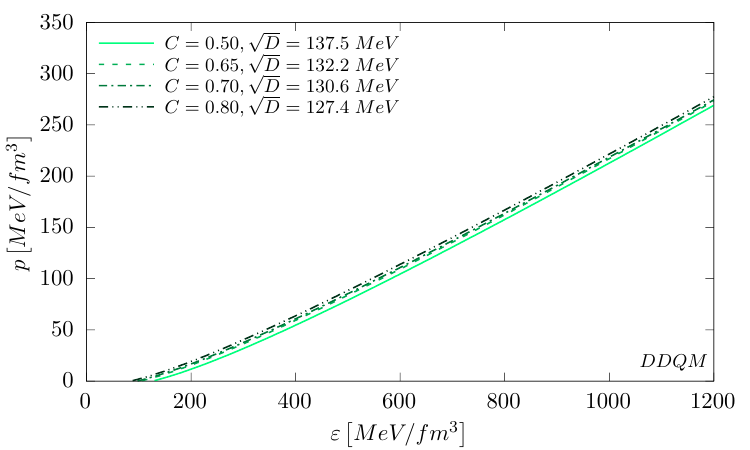}
     \end{subfigure}
     \begin{subfigure}{0.49\textwidth}
         \centering
         \includegraphics[width=\textwidth]{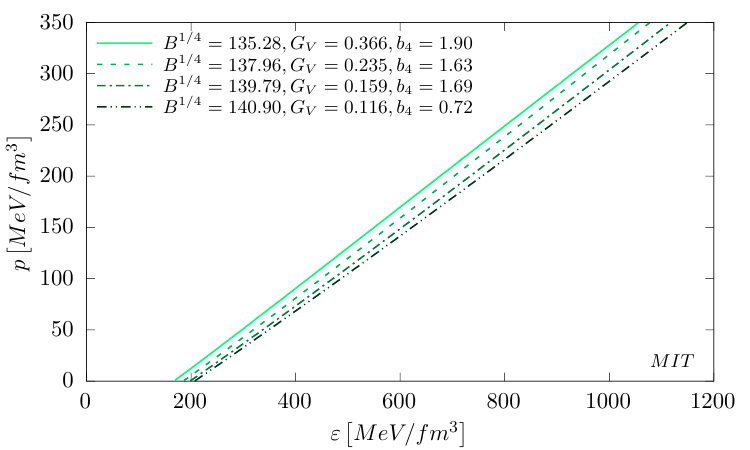}
     \end{subfigure}
        \caption{In the figures above we compare the EoSs from the DDQM (left panel) with the EoSs for the vector MIT bag model (right panel).}  
        \label{fig9}
\end{figure} 

\begin{table}[t]
    \centering
        \caption{Neutron star properties for the different DDQM models analyzed}
    \begin{ruledtabular}
    \begin{tabular}{c c c c c c c c}
         $C$ & $\sqrt{D}[{\rm MeV}]$ & M$_{max}[{\rm M}_{\odot}]$ & $R[{\rm km}]$ & $n[{\rm fm}^{-3}]$ & $R_{1.4}[{\rm km}]$ & $\Lambda_{1.4}$\\
         \hline
        0.50  & 137.5  & 1.91  & 11.78  & 0.88 & 12.46  & 534  \\

        0.65  & 132.2  & 2.04  & 12.82  & 0.73 & 13.40  & 1398  \\

        0.70  & 130.6  & 2.10  & 13.25  & 0.70 & 13.86  & 1717 \\ 

        0.80  & 127.4  & 2.18  & 13.86  & 0.64 & 14.41  & 2163
    \end{tabular}
    \end{ruledtabular}
    \label{tabdd}
\end{table}

\begin{table}[t]
    \centering
        \caption{Neutron star properties for the different vector MIT bag model analyzed}
    \begin{ruledtabular}
    \begin{tabular}{c c c c c c c c}
         $B^{1/4}[{\rm MeV}]$  &  $G_V[{\rm fm}^2]$  &  $b_4$ & M$_{max}[{\rm M}_{\odot}]$ & $R[{\rm km}]$ & $n[{\rm fm}^{-3}]$ & $R_{1.4}[{\rm km}]$ & $\Lambda_{1.4}$\\
         \hline
          135.28 & 0.366 & 1.90 & 2.54 & 13.15 & 0.74 & 12.38 &  1078 \\

          137.96 & 0.235 & 1.63 & 2.40 & 12.46 & 0.82 & 11.94 & 850 \\

          139.79 & 0.159 & 1.69 & 2.28 & 11.96 & 0.89 & 11.63 & 712 \\ 

          140.90 & 0.116 & 0.72 & 2.21 & 11.66 & 0.94 & 11.43 & 633
    \end{tabular}
    \end{ruledtabular}
    \label{tabMIT}
\end{table}

\begin{figure}[t]
     \centering
     \begin{subfigure}{0.48\textwidth}
         \centering
         \includegraphics[width=\textwidth]{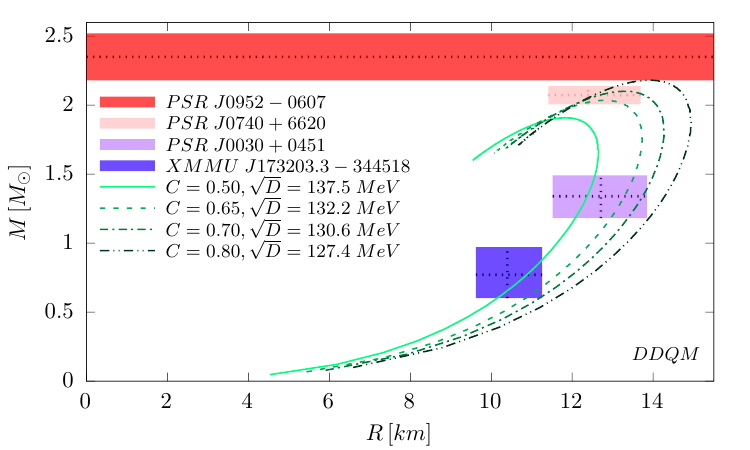}
     \end{subfigure}
     \begin{subfigure}{0.48\textwidth}
         \centering
         \includegraphics[width=\textwidth]{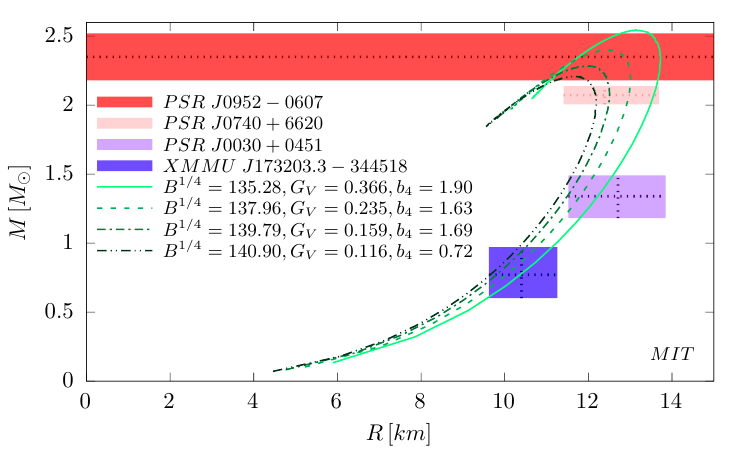}
     \end{subfigure}
        \caption{Comparing the mass-radius diagram of the DDQM (left) and the vector MIT bag model(right).}
        \label{fig6}
\end{figure}

\begin{figure}[t]
     \centering
     \begin{subfigure}{0.49\textwidth}
         \centering
         \includegraphics[width=\textwidth]{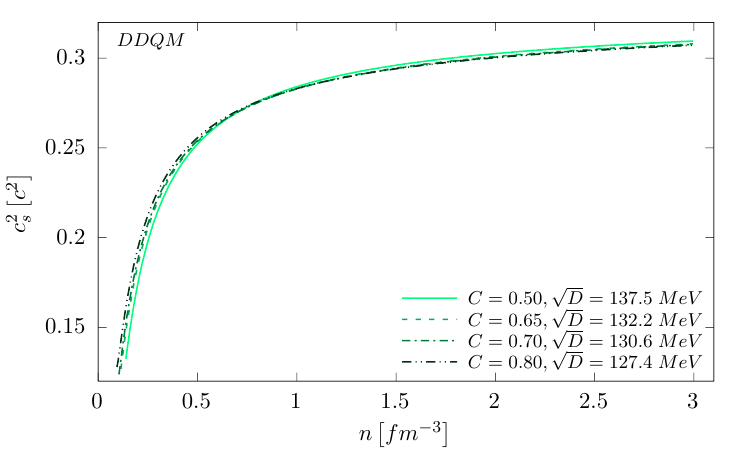}
     \end{subfigure}
     \begin{subfigure}{0.49\textwidth}
         \centering
         \includegraphics[width=\textwidth]{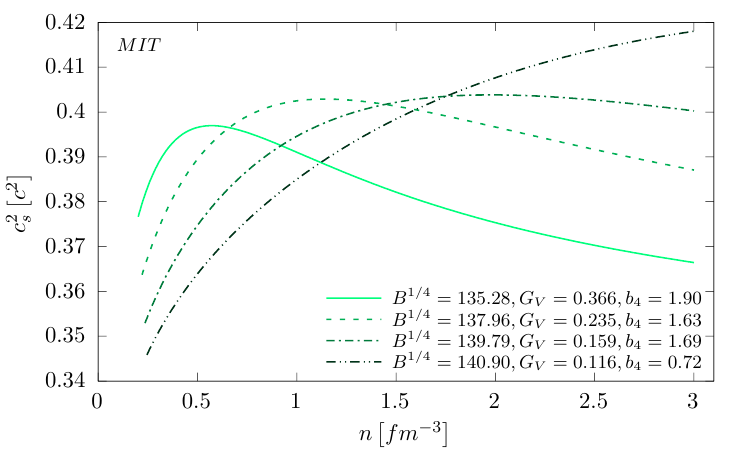}
     \end{subfigure}
        \caption{Here we compare the square velocity of sound $c_s^2$ resulting from the two models as a function of baryon density $n$.}
        \label{fig7}
\end{figure} 

\begin{figure}[t]
     \centering
     \begin{subfigure}{0.49\textwidth}
         \centering
         \includegraphics[width=\textwidth]{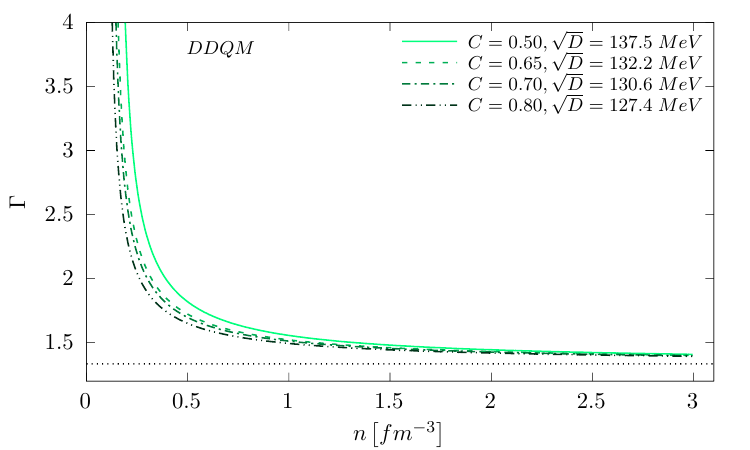}
     \end{subfigure}
     \begin{subfigure}{0.49\textwidth}
         \centering
         \includegraphics[width=\textwidth]{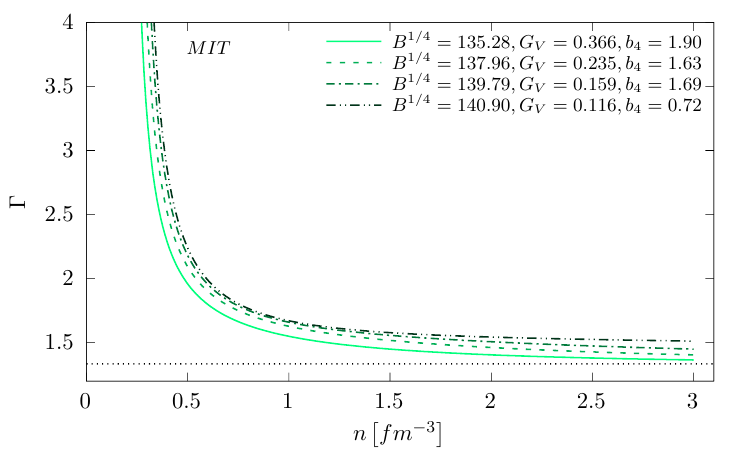}
     \end{subfigure}
        \caption{We compare the adiabatic indices $\Gamma$ of DDQM (left panel) and the vector MIT bag model (right panel) as a function of baryon density $n$.}
        \label{fig8}
\end{figure} 

\begin{figure}[t]
     \centering
     \begin{subfigure}{0.49\textwidth}
         \centering
         \includegraphics[width=\textwidth]{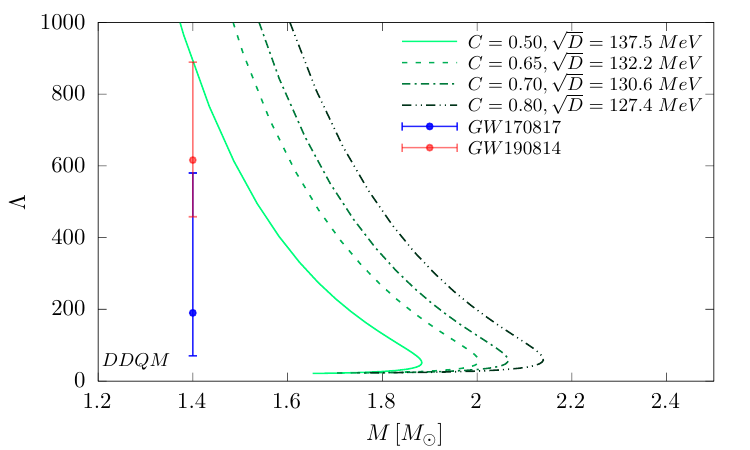}
     \end{subfigure}
     \begin{subfigure}{0.49\textwidth}
         \centering
         \includegraphics[width=\textwidth]{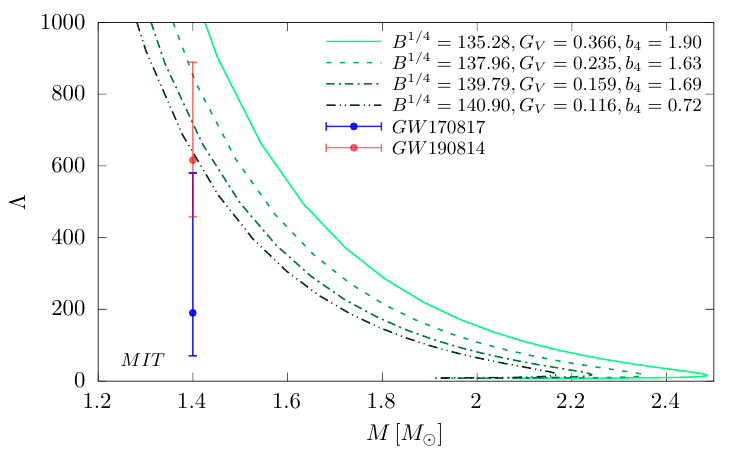}
     \end{subfigure}
        \caption{Tidal deformability $\Lambda$ as a function of the mass $M[M_{\odot}]$ for the DDQM model (left panel) and vector MIT bag model (right panel).}
         \label{fig10}
\end{figure}

Since the models currently under consideration are associated with various constants, see Tables~\ref{tabdd} and \ref{tabMIT}, we resort to Bayesian analysis, shown in Figures~\ref{fig1}, \ref{fig2}, \ref{fig3} and \ref{fig4}, to determine the values of these constants that are most consistent with the constraints of some selected stars presented on Table~\ref{tab1}. For instance, the vector MIT bag model is associated with three free parameters, although they are not fully independent from each other due to the stability window~\cite{Lopes_2021}, whilst the DDQM, is associated with two free parameters. In Table~\ref{tabdd} we display the results for $C$ and $\sqrt{D}$ related to the DDQM  and in Table~\ref{tabMIT} we display the results for $B^{1/4},\;G_V\; \text{and}\; b_4$ determined from the MIT bag model. Even though we impose the same constraints on the two models, we observe that the results obtained for the vector MIT bag model lead to stiffer EoSs than for the DDQM model. This, in turn, leads to higher maximum masses and smaller radii in the vector MIT bag case, as well as lower tidal deformability values when compared to the DDQM case, as will be evident in the discussions that follow.

In Figure~\ref{fig9}, we present the EoSs for both the DDQM model and the vector MIT bag model composed of three flavor quarks ($u, d, s$) and electrons calculated in $\beta$-equilibrium. The stiffness of the EoS enables us to determine the maximum mass of the star. From Figure~\ref{fig9}, right panel, we observe that an increase in the $B^{1/4}$ and/or a decreasing $G_V$ softens the EoS. In the same sense, an increase of  $b_4$ leads to the softening of the EoS at high densities. Looking at Table \ref{tabMIT} and Figure \ref{fig9} one might be led to believe that increasing $b_4$ leads to a stiffer EoS but, if we keep the other parameters fixed and vary only $b_4$ we find that it effect is softening the EoS, as can be seen in \cite{Lopes_2021}. What happens in the present case is that  $B^{1/4}$ and $G_V$ have a greater influence on the EoS than $b_4$. 

On the other hand, on the left panel, we observe that an increasing $C$ and associated decreasing $\sqrt{D}$ leads to the stiffening of the EoS, resulting in increasing maximum masses in that order, as can be seen in Table~\ref{tabdd}. These properties have a direct impact on the velocity of sound $c_s$, the adiabatic index $\Gamma$, and the tidal deformability $\Lambda$ that enables us to study the inner composition of the star. 

As it is well known, to satisfy the $2$ M$_\odot$ constraint of NSs, stiffer EoSs are the ones preferred. The results also show that the  EoS for the vector MIT bag model is significantly stiffer when compared to the DDQM. Since stiffer EoS means higher maximum masses, this is reflected in the maximum masses calculated in the framework of the models. While all the curves for the vector MIT bag model satisfy the constraint of having a maximum mass compatible with the estimated mass of PSR J0952-0607, only one of the curves -- dash double-dot curve --  satisfies this constraint for the DDQM model. The  two other curves -- dashed curve and dash-dotted curve -- still present a maximum mass compatible with the estimated mass for PSR J0740+6620.

In Figure \ref{fig6}, we show the mass-radius diagram for the DDQM model on the left panel and for the vector MIT bag model on the right panel. Starting with the left panel for the DDQM model, we can observe that the softer EoS, solid curve, only satisfies the constraints for the XMMU J173203.3-344518 object and the  PSR J0030 + 0451 pulsar
and does not achieve M$_{max}=2$ M$_{\odot}$. The two curves that came from the EoSs with intermediate stiffness for this model satisfy the constraints from NICER and the one from XMMU J173203.3-344518. As for the curve with stiffer EoS, it is the only one for the DDQM case that satisfies the mass of the ``black widow'' star with a radius of $13.86$ km. For this model, we are not able to find a set of values for the parameters $C$ and $\sqrt{D}$ that would satisfy all the constraints at the same time. Now, analyzing the second panel of Figure \ref{fig6} for the vector MIT bag model, we can see that all the curves can satisfy all the mass-radius constraints at the same time. Despite that, the dashed double-dotted curve which is associated with the softest EoS of this model, only satisfies the constraints for the canonical and the ``black widow'' stars slightly. We can also see that the curve with stiffer EoS for the MIT bag case, the solid curve, has a maximum mass above the estimated range of mass for PSR J0952-0607, M$_{max} = 2.54$ M$_{\odot}$. In general, comparing the two models we can infer that, for the cases we chose to study, i.e., 
after the parameters are restricted to the ones suggested by the Bayesian calculation,
the DDQM model leads to smaller maximum masses and higher radii than the vector MIT bag model.

The QCD theory shows different properties in the perturbative and nonperturbative regions. Quark matter is expected to be in a deconfined state and approximately symmetric under conformal transformation. On the other hand, the hadronic matter is not symmetric under conformal transformation due to the manifestation of chiral symmetry breaking. These two extreme characteristics can be quantitatively differentiated through the determination of the speed of sound, $c_s^2=dp/d\varepsilon$ in the stellar matter.  It has been established that $c^2_s$ is constant and attains a value of $1/3$ in exactly conformal matter and approaches this value from below at high-density quark matter region,  $n>40~n_0$ \cite{PhysRevD.81.105021}. Hence, we can determine the interior dynamics and composition of the star by analyzing the $c_s$, which depends on the EoS of the corresponding star. The $c_s$ is determined to be $c_s^2 \ll 1/3$ below the saturation density in chiral effective theory and can grow up to $c_s^2 \gtrsim 0.5$ in hadronic matter at high densities \cite{Bedaque:2014sqa, Annala:2019puf}. Causality requires that $c_s^2\leq 1$ and thermodynamic stability also requires that $c_s^2 > 0$. However, if the interaction between the particles is perturbative, $c_s^2 \leq 1/3$. This is applicable to the case of QCD at asymptotically high density or temperature where perturbative treatment of the theory is valid. 

Comparing the graphs in Figure~\ref{fig7}, in the left panel for the DDQM model, all the curves lie below the conformal limit $c^2_s < 1/3$. This implies that, for the possible values of $C$ and $\sqrt{D}$ obtained for the constraints considered in this work, the stars can possibly be formed through self-bound free quarks in a deconfined state. On the other hand, the vector MIT bag model on the right panel shows characteristics similar to the chiral effective theory, that is, the curves show $c_s^2 > 1/3$ even at very low densities. Aside from that, the curves with higher values of $b_4$, the solid curve and dashed one, show different characteristics from the other two, they rise steadily at low density and start falling at intermediate to high densities. Similar characteristics are observed among hybrid neutron stars, where deconfined quark matter is assumed to be in the core of the stars \cite{10.1093/mnras/stad2509, PhysRevC.105.045808, Xia:2019xax, Kojo:2020krb}. In general, in the framework of the vector MIT bag model, the $c_s^2$ behaves analogous to the hadronic matter at the lower to intermediate densities and then drops close to the conformal limit in the core of the star where the quark core is assumed to have formed. The other two curves, the dash-dot curve and the dash double-dot one, follow the usual characteristics of the $c_s^2$ in the hadronic matter, where $c_s^2$ rises steadily with $n$ but exceeds the conformal limit $c_s^2 = 1/3$ \cite{Altiparmak:2022bke}.  The decrease of $c_s^2$ with the density shows the importance of the self-coupling that mimics the Dirac sea contribution, as the conformal limit must be satisfied at very high densities.

In Figure~\ref{fig8}, we analyze the stability of the stars using the adiabatic index as a benchmark. Following the seminal works of Chandrasekhar \cite{PhysRevLett.12.114, Chandrasekhar:1964zz}, the dynamic stability of a star can be analyzed based on variational methods.  An expression for the adiabatic index is given by
\begin{equation}
    \Gamma = \dfrac{p+\varepsilon}{p}\left(\dfrac{d p}{d\varepsilon}\right)_S,
\end{equation}
where $dp/d\varepsilon$ is precisely the speed of sound and $S$ is the specific entropy at which $\Gamma$ is evaluated. Generally, the $\Gamma$ is dimensionless and its behavior depends on the stiffness of the EoS for spherical relativistic fluid. For a stable star, the adiabatic index is required to be $\Gamma > \Gamma_{cr}=4/3$ in the core of the star. Meanwhile, collapse of the star is expected to begin when $\Gamma$ falls below $4/3$, $\Gamma_{cr}$ is the critical adiabatic index \cite{glass1983stability, 2022EPJC...82...57T, Haensel:1986qb, PhysRevC.50.460}. The case of $\Gamma = 4/3$ is the starting point of instability of the star. In Figure~\ref{fig8}, we demarcate the instability threshold $\Gamma_{cr}$ with a dotted horizontal gray line. We observe that the $\Gamma$ decreases with increasing $n$ 
but does not cross the instability threshold for both models under consideration. Additionally, we observe that more massive stars with stiffer EoSs approach $\Gamma_{cr}$ line faster than relatively lighter stars with softer EoSs.

The NS macroscopic properties such as the masses and radii have long been used as constraints to understanding the microscopic properties of these stars. Despite the extensive probe of the NSs, some of its key properties and interior compositions at extreme conditions of density and isospin asymmetry still remain uncertain. Here, we analyze another astrophysical observable property, the tidal deformability presented in Figure~\ref{fig10}, that can also be used to probe the interior composition of the NS. The NSs like any other external objects with a defined structure can tidally deform when subject to the influence of an external tidal field. During the event of coalescences of NSs that led to gravitational wave emission, the deformation was quantified through a dimensionless parameter, called the tidal deformability $\Lambda$. The $\Lambda$ is given by the expression \cite{lopes2022nature,Chatziioannou:2020pqz,Flores2020,Lourenço2021}: 
\begin{equation}
    \Lambda =\dfrac{2}{3}k_2\dfrac{R^5}{M^5},
\end{equation}
where $k_2$ is the gravitational Love number, $M$ is the mass of the star and $R$ is its radius. As expected, a relatively larger value of $\Lambda$ implies the star is large, less compact, and can easily be deformed. On the contrary, a smaller $\Lambda$ means, a smaller-sized star, highly compact, and hard to deform.  Moreover, as pointed in references \cite{lopes2022nature,Lourenço2021}, the value of $y_R$  must be corrected, since strange stars are self-bound and present a discontinuity at the surface. Therefore we must have:
\begin{equation}
 y_R \rightarrow y_R - \frac{4\pi R^3 \Delta\varepsilon_S}{M} , \label{EL26}
\end{equation}
where $R$ and $M$ are the star radius and mass respectively, and $\Delta\varepsilon_S$ is the difference between the energy density at the surface ($p =0$) and the exterior of the star (which implies $\varepsilon=0$).

We considered two events that satisfy some of our results in Figure~\ref{fig10} for both the DDQM and the vector MIT bag model. We first analyze event GW170817, which is arguably the most authoritative confirmed observed binary neutron star merger with an emitted gravitational wave as of now \cite{LIGOScientific:2017vwq, LIGOScientific:2018hze}. The value of $\Lambda_{1.4}$ estimated for this event is $\Lambda_{1.4} = 190^{+390}_{-120}$ at $90\%$ confidence level \cite{LIGOScientific:2017vwq}. If the DDQM model is used, this constraint is not satisfied by any of the EoSs analysed. As for the MIT bag model, none of its curves satisfy this constraint either. We also demarcated the binary coalescence event GW190814, believed to consist of a black hole of mass within $22.2 - 24.3$ M$_{\odot}$ and a compact object with a mass within $2.50 - 2.67\, {\rm M}_\odot$ \cite{LIGOScientific:2020zkf}. Analysis of the data associated with GW190814 shows that the possible nature of the compact object can be classified as a neutron star only if its EoS is very stiff \cite{Huang:2020cab} or it is a rapidly rotating compact object below the mass shedding frequency \cite{Zhang:2020zsc, PhysRevLett.125.261104, PhysRevC.102.065805}. Its value has been estimated to be $\Lambda_{1.4} = 616^{+273}_{-158}$ \cite{LIGOScientific:2020zkf}. This constraint is satisfied only by the solid curve in the DDQM model. In the MIT bag model case, the three curves associated with the three softest EoSs satisfy this constraint. Hierarchically, curves with lower masses are the first to satisfy the constraints on tidal deformability before the massive ones in both model frameworks. The absolute values of $\Lambda_{1.4}$ determined in the two separate models analyzed here are presented on Tables~\ref{tabdd} and \ref{tabMIT}, where we can see that the value of $\Lambda_{1.4}$ decreases with the increasing of the parameter $C$ and decreasing of $\sqrt{D}$ for the DDQM model. As for the MIT bag model, the absolute value of $\Lambda_{1.4}$ decreases with increasing $B^{1/4}$ and decreasing $G_V$ as can be seen in Table \ref{tabMIT}. From Table \ref{tabMIT}, it seems that increasing $b_4$ leads to increasing $\Lambda_{1.4}$ as well but, since we know that increasing $b_4$ softens the EoS \cite{Lopes_2021}, then increasing $b_4$ will decrease $\Lambda_{1.4}$. In general, the values of the tidal deformability for the MIT bag case are smaller than the values found for the DDQM case.

\section{Conclusions} \label{conclusions}

In this paper, we perform a comparative analysis between the DDQM and the vector MIT bag model and their applications to the study of quark stars that satisfy the constraints of some observed pulsars and compact objects listed in Table~\ref{tab1} using Bayesian analysis. We show the corner plot for the distribution of the various parameters determined at various confidence levels in Figures~\ref{fig1}, \ref{fig2}, \ref{fig3} and \ref{fig4} for Cases I, II III and IV respectively. We imposed four different mass and radius constraints (corresponding to four different compact objects) on the EoSs for each model and determined the model parameters that satisfy these constraints. The parameters determined through this analysis for the vector MIT bag and the DDQM models are presented in Tables~\ref{tabMIT} and \ref{tabdd} respectively. In the case of the vector MIT bag model, we were able to include the stability window analysis in the Bayesian inference, so that, all points in the corner plots for this model are inside the stability window. In the case of the DDQM, we performed a stability window analysis separately, as can be seen in Figure~\ref{fig5}, to clearly show the regions of $C$ (associated with the leading-order perturbative term in QCD) and $\sqrt{D}$ (associated with linear confinement) within which the corresponding EoS will lead to the determination of a stable quark star, according to the Bodmer-Witten conjecture.

After obtaining the model parameters from the analysis, we applied them to study the properties of quark stars assuming they have the same constraints as NSs. Consequently, we calculated the EoSs for the two models under investigation for each set of values of the parameters that were chosen to be analyzed and determined their speed of sound ($c_s^2$),  adiabatic index ($\Gamma$), tidal deformability ($\Lambda$) and mass-radius diagram (M-R) and compared their similarities and differences.
\begin{itemize}
    
    \item We find that the EoSs determined from the vector MIT bag model are stiffer than the ones determined from the DDQM even though we imposed similar constraints on both models, see the result in Figure~\ref{fig9}. In this case, we can infer that different models can show different dynamics of EoS even if the same constraints are used in calculating them. Therefore, we expect stars with similar constraints to show different characteristics in each model framework.
    \item In Figure~\ref{fig6}, we present the mass-radius diagram obtained from both models. In the framework of the vector MIT model, all the stars satisfied the $2$ M$_\odot$ mass constraint imposed on NSs. As mentioned above, the EoS determined in this model are stiffer compared to DDQM, hence the higher masses are expected. On the other hand, the maximum masses of the stars within the DDQM also satisfy the $2$ M$_\odot$ constraint except one, the solid line curve with M$_{\rm max} = 1.91$ M$_\odot$.
    \item In Figure~\ref{fig7}, we present the $c_s^2$ as a function of $n$ for both models to help determine the inner composition of the star. Here, the DDQM showed a characteristic similar to deconfined quark matter with $c^2\,<\,1/3$. On the other hand, the vector MIT bag model crossed the conformal limit at $c^2=1/3$ even at the low-density regions showing a behavior similar to hadron matter. Also, the curves that have higher maximum masses for this model (solid and dashed line curves in the right panel) showed a steady rise in $c_s$ with $n$ at low-density regions and started falling at intermediate to higher-density regions. 
    
    \item  In Figure~\ref{fig8}, we analyze the stability of the stars in both models through their adiabatic indices. Generally, all the stars analyzed are well within the stability threshold $\Gamma\,>\,\Gamma_{cr}$. However, we observed that curves associated with more massive stars approach the $\Gamma_{cr}$ line (the gray line) faster than the curves related to smaller M$_{max}$ in both models. 
    
    \item  Additionally, we studied the tidal deformability that complements the study of the interior dynamics of the NSs. Generally, we observed that curves with higher maximum mass, which at the same time have canonical stars with higher radius, have higher values of $\Lambda_{1.4}$ than the curves with lower M$_{max}$ and more compact canonical stars, for the same model. The stars determined in the framework of the vector MIT bag model have larger maximum masses, smaller radii, and smaller values of $\Lambda_{1.4}$ compared to the ones determined from the DDQM with relatively low masses and higher radii. As expected, less compact stars are more likely to be deformed than the more compact ones in the same model framework. Only one of the curves (solid line) in the DDQM falls in the upper limit of the constraints imposed by GW 190814 and three of the curves (dash double-dotted, dash-dotted, and dashed lines) in the MIT bag model satisfy the GW190814 constraint as well, as can be seen in Figure~\ref{fig10}.

\end{itemize}
Consequently, the optimized model parameters for the MIT bag and the DDQM models determined through the analysis qualitatively reproduce some known NS properties such as the one listed above. Most of the results obtained conform with the $2$ M$_\odot$ maximum constraint imposed on NSs \cite{riley2021nicer}. Some of the results obtained for the $\Lambda_{1.4}$ satisfy the GW170817 \cite{LIGOScientific:2018hze, LIGOScientific:2017vwq} (DDQM) and GW190814 \cite{LIGOScientific:2020zkf, Huang:2020cab} (MIT bag model) signal ranges. 

\section*{Acknowledgements}

This work is a part of the project INCT-FNA Proc. No. 464898/2014-5. D.P.M. was partially supported by Conselho Nacional de Desenvolvimento Científico e Tecnológico (CNPq/Brazil) under grant 303490-2021-7 and A.I. under grant 168546/2021-3. F.M.S. would like to thank FAPESC/CNPq for financial support under grant 150721/2023-4. L.C.N.S would like to thank CNPq for partial financial support through the research Project No. 200215/2023-0. LLL was partially supported by CNPq Universal Grant No. 409029/2021-1.

\bibliographystyle{ieeetr}
\bibliography{ref.bib}

\end{document}